\documentclass[
 amsmath,amssymb,
 aip,
jcp,onecolumn,11pt]{revtex4-2}

\usepackage{graphicx}
\usepackage{dcolumn}
\usepackage{bm}
\usepackage{color}
\usepackage{xcolor}

\newcommand{\BgradB}{$(\mathbf{B} \cdot \nabla) \mathbf{B}$}
\newcommand{\Peclet}{P{\'e}clet }

\newcommand{\modified}[1]{{\color{black}{#1}}}

\begin{document}

\title{Magnetophoresis of paramagnetic nanoparticles in suspensions under magnetic field gradients}

\author{Peter Rassolov}
\affiliation{Department of Chemical and Biomedical Engineering, FAMU-FSU College of Engineering, Tallahassee, FL, 32310, USA}
\affiliation{Center for Rare Earths, Critical Minerals, and Industrial Byproducts, National High Magnetic Field Laboratory, Tallahassee, FL 32310, USA}
\author{Jamel Ali}
\affiliation{Department of Chemical and Biomedical Engineering, FAMU-FSU College of Engineering, Tallahassee, FL, 32310, USA}
\affiliation{Center for Rare Earths, Critical Minerals, and Industrial Byproducts, National High Magnetic Field Laboratory, Tallahassee, FL 32310, USA}
\author{Theo Siegrist}
\affiliation{Department of Chemical and Biomedical Engineering, FAMU-FSU College of Engineering, Tallahassee, FL, 32310, USA}
\affiliation{Center for Rare Earths, Critical Minerals, and Industrial Byproducts, National High Magnetic Field Laboratory, Tallahassee, FL 32310, USA}
\author{Munir Humayun}

\affiliation{Department of Earth, Ocean and Atmospheric Science, Florida State University, Tallahassee, FL 32304, USA.}
\affiliation{Center for Rare Earths, Critical Minerals, and Industrial Byproducts, National High Magnetic Field Laboratory, Tallahassee, FL 32310, USA}
\author{Hadi Mohammadigoushki}
\thanks{Corresponding author}\email{hadi.moham@eng.famu.fsu.edu}
\affiliation{Department of Chemical and Biomedical Engineering, FAMU-FSU College of Engineering, Tallahassee, FL, 32310, USA}
\affiliation{Center for Rare Earths, Critical Minerals, and Industrial Byproducts, National High Magnetic Field Laboratory, Tallahassee, FL 32310, USA}

\date{\today}

\begin{abstract}

We systematically investigate the magnetophoresis of weakly paramagnetic manganese oxide nanoparticles under nonuniform magnetic fields using a combination of experiments and multiphysics numerical simulations. Experiments were conducted in a closed cuvette exposed to a nonuniform magnetic field generated by an electromagnet, covering a wide range of particle concentrations (25–200 mg/L) and magnetic field gradients (0–110 T$^2$/m). The experimental results reveal that paramagnetic manganese oxide nanoparticles exhibit significant magnetophoretic behavior, leading to particle depletion within the cuvette. The depletion rate is independent of the initial particle concentration but strongly depends on the magnetic field gradient. At low magnetic field gradients, magnetophoresis progresses slowly, while at higher gradients, the particle depletion rate increases significantly before stabilizing. Transient concentration gradients emerge within the cuvette during magnetophoresis, which we hypothesize are driven by magnetic Grashof numbers ($\mathrm{Gr}_m$ near unity. When $\mathrm{Gr}_m >1$, the formation of concentration gradients induces bulk fluid flows that accelerate particle capture at regions of maximum magnetic field strength. In systems where magnetophoresis opposes sedimentation, particle-depleted regions form when the ratio of magnetic to gravitational \Peclet numbers exceeds 1. The numerical simulations suggest formation field-induced aggregation for manganese oxide nanoparticles with radii of 130 nm or larger. These insights highlight the potential of magnetic separation for sustainable metal recovery, offering a scalable and environmental friendly solution for recycling critical materials from spent electronics.
\end{abstract}

\maketitle
\section{Introduction}

Magnetophoresis, or the transport of particles driven by differences in magnetization from the surrounding media under nonuniform magnetic fields, presents a promising mechanism for controlling the movement of magnetic materials with applications in drug delivery~\cite{murthy2010magnetophoresis,alexander2012approaches}, water purification~\cite{ambashta2010water}, and chemical separation~\cite{pamme2004chip,benhal2025dynamics}. Magnetophoresis offers several advantages over traditional separation techniques, including high selectivity, energy efficiency, rapid separation, non-invasiveness, and scalability \cite{benhal2025dynamics,pamme2004chip,han2004continuous}. When a magnetic particle suspended in a fluid medium is exposed to an external magnetic field, the total magnetic energy that it experiences is given by~\cite{john1967foundations}: 
\begin{equation}\label{eqn1}
    E_m=-\frac{1}{2\mu_0}\Delta\chi_m c B^2
\end{equation}
where $\Delta\chi_m$ is the difference between molar magnetic susceptibility of the particle and the surrounding medium, \textbf{B} is the magnetic flux density, $c$ is the particle concentration, and $\mu_0$ is the vacuum permeability ($4 \pi \times 10^{-7}$ N/A$^2$). Under isothermal conditions, and applying the first law of thermodynamics, the magnetic force on the particle is proportional to the size of the particle, the difference in magnetic susceptibility of the particle from the surrounding medium, and the gradient of the magnetic field. Depending on the type of the system (closed vs. open), the magnetophoretic force must compete with various opposing forces to induce net magnetophoresis. These competing forces may include random thermal fluctuations, gravity, inertia, or hydrodynamic forces resulting from bulk fluid flow. \par 

The literature on magnetophoresis of nanometer-sized and larger particles is quite expansive, ranging from high gradient magnetic separation (HGMS) of micron and/or nanoparticles of strongly magnetic particles\cite{ge2017magnetic,watson1975theory,watson1973magnetic,iranmanesh2017magnetic,ditsch2005high,moeser2004high,Svoboda2004} to low gradient magnetic field separation (LGMS) of superparamagnetic particles based on iron oxide\cite{Leong2020Langmuir,Leong2015SoftMatter,yavuz2006low,lim2014challenges,schaller2008motion,de2008low}. Under certain conditions, the experimentally observed rate of particle removal in LGMS significantly exceeds the rate predicted for particles moving independently through a stagnant fluid medium \cite{yavuz2006low}. This unexpected result was hypothesized to arise from the formation of field-induced nanoparticle clusters, where applied magnetic fields drive the reversible aggregation of particles \cite{yavuz2006low}. This observation spurred a new wave of studies that not only provided direct evidence for the formation of nanoparticle clusters but also offered an in-depth understanding of the mechanisms driving magnetic field-induced cluster formation~\cite{klokkenburg2006quantitative,klokkenburg2006situ,socoliuc2013magnetically,laskar2010experimental,gajula2010visualization,ezzaier2017two,Andreu2011aggregation,faraudo2013understanding}.\par 
While considerable progress has been made in understanding the magnetically assisted separation of superparamagnetic fine particles (micro- and nano-sized), a notable gap still exists in determining the method's effectiveness for separating weakly paramagnetic materials, which have magnetic susceptibilities much smaller (by orders of magnitude) than the superparamagnetic particles. In several important applications, weakly paramagnetic materials are present in complex mixtures and need to be separated. For example, valuable metals such as cobalt, manganese, and nickel, found in lithium-ion batteries (LIBs), are essential for their functionality \cite{chan2021closed,whittingham2004lithium}. Similarly, various electronic devices including semiconductors and LEDs commonly contain precious and critical metals such as silver, gold, gallium, and manganese among other weakly paramagnetic metals that differ in their magnetic susceptibility\cite{cenci2022precious}. With the growing production of LIBs and the increasing demand for electronics, millions of electronic devices will reach the end of their lifespan; efficient recycling of the metals contained in them would reduce demand for natural resources, enhance supply chains and ameliorate environmental concerns arising from disposal of discarded devices~\cite{icsildar2019biotechnological,chan2021closed}. Magnetic separation offers a promising method to efficiently separate and recover these precious metals from spent electronics.\par 

The primary goal of this paper is to evaluate the magnetophoresis of weakly paramagnetic materials in a non-uniform magnetic field. For this purpose, we utilize manganese oxide-based nanoparticles, which exhibit one of the strongest magnetic susceptibilities among transition metals. Given that the magnetic force acting on weakly paramagnetic nanoparticles is relatively small, it must overcome thermal fluctuations to induce net magnetophoresis. We examine the effects of varying magnetic field strength, initial particle concentrations, and the interplay between magnetic forces and gravity. Additionally, we explore whether the formation of field-induced clusters or secondary convective flows, resulting from particle-fluid interactions, could enhance the magnetophoresis process. Experiments are conducted in a closed cuvette exposed to a non-uniform magnetic field generated by an electromagnet. The experiments are complemented by multiphysics numerical simulations to provide a deeper understanding of the underlying mechanisms of magnetophoresis and the factors influencing the behavior of weakly paramagnetic materials in non-uniform magnetic fields.



\section{Theory}

\subsection{Magnetic field}

In this work, the magnetic field is steady throughout the duration of an experiment. The magnetic field is given by Ampere's law, and the constitutive relation for the magnetic flux density is assumed to be a known relative permeability:
\begin{subequations}
\begin{align}
\nabla \times \mathbf{H} & = \mathbf{J}_e \label{eq:Amperes-law}\\
\mathbf{B} & = \nabla \times \mathbf{A} \label{eq:magnetic-potential} \\
\mathbf{B} & = \mu_0 \mu_r \mathbf{H}, \label{eq:relative-permeability}
\end{align}
\end{subequations}
where $\mathbf{H}$ is the magnetic field strength, $\mathbf{J}_e$ is the current density, $\mathbf{B}$ is the magnetic flux density, $\mathbf{A}$ is the magnetic vector potential, and $\mu_r$ is the relative permeability. Of these, $\mathbf{J}_e$ is assumed to be zero outside the coils and is held fixed at the applied current density in each coil. The relative permeability $\mu_r$ is close to unity except in the magnet poles, where it is obtained from the magnetization curve for soft iron in the COMSOL materials library.

\subsection{Suspended species transport}
Under a nonuniform external magnetic field, a spherical particle suspended in a fluid experiences a net magnetic force given by the following~\cite{boyer1988force}:
\begin{equation}
\mathbf{F}_m = \frac{4 \pi}{3} R_p^3 \frac{\Delta \chi_V}{\mu_0} \left( \mathbf{B} \cdot \nabla \right) \mathbf{B},
\end{equation}
where $R_p$ is the radius of the magnetizable volume of the particle, $\Delta \chi_V$ is the difference in magnetic susceptibility between the particle and the surrounding medium, and $\mathbf{B}$ is the magnetic flux density. The magnetic force induces movement of the particle, known as magnetophoresis. In the absence of inertia, the resulting motion is opposed by drag from the fluid medium that is given by Stokes's law:
\begin{equation}
\mathbf{F}_d = -6 \pi \eta \mathbf{v}_m R_h,
\label{eq:Stokes-drag}
\end{equation}
where $\eta$ is the dynamic viscosity of the fluid, $\mathbf{v}_m$ is the steady-state drift velocity due to magnetophoresis, and $R_h$ is the hydrodynamic radius of the particle (which may be larger than $R_p$). A force balance ($\mathbf{F}_m + \mathbf{F}_d = \mathbf{0}$) is applied to obtain $\mathbf{v}_m$:
\begin{equation}
\mathbf{v}_m = \frac{2 R_p^3 \Delta \chi_V}{9 \mu_0 \eta R_h} \left( \mathbf{B} \cdot \nabla \right) \mathbf{B}
\label{eq:magph-velocity}
\end{equation}

In addition to magnetophoresis, particles undergo diffusion due to random thermal motions. The diffusion gives rise to the diffusive flux $\mathbf{J}$:
\begin{equation}
\mathbf{J} = -D \nabla c
\label{eq:Ficks-law}
\end{equation}
where $D$ is the diffusion coefficient and $c$ is the concentration of the suspended species (mass of the particle material per unit volume suspension). For particles much larger than the size of the fluid molecules, the Stokes-Einstein relation can be used to obtain the diffusion coefficient~\cite{Leong2015SoftMatter}:
\begin{equation}
D = \frac{k_B T}{6 \pi \eta R_h}
\end{equation}
where $k_B$ is the Boltzmann constant and $T$ is the temperature.

Third, gravity acts on the particles and causes sedimentation. The force of gravity on a particle is given by:
\begin{equation}
\mathbf{F}_g = \frac{4 \pi}{3} R_p^3 \Delta \rho \mathbf{g}
\end{equation}
where $\Delta \rho$ is the difference in density between the particle and the surrounding medium and $\mathbf{g}$ is the free-fall acceleration due to gravity (9.81 m/s$^2$ downward). This is also opposed by drag given by Stokes law (Eq.~\ref{eq:Stokes-drag}) with a resulting velocity:
\begin{equation}
\mathbf{v}_g = \frac{2 R_p^3 \Delta \rho}{9\eta R_h} \mathbf{g}
\label{eq:sed-velocity}
\end{equation}

Finally, the fluid medium in the system can flow as a bulk. The bulk fluid flow requires solving the Navier-Stokes equations for the flow velocity $\mathbf{v}_f$ and is discussed in the following section.
Equations~\ref{eq:magph-velocity},~\ref{eq:Ficks-law}, and~\ref{eq:sed-velocity}, can be combined with $\mathbf{v}_f$ to obtain the total flux for suspended nanoparticles:
\begin{equation}
\mathbf{N} = -\frac{k_B T}{6 \pi \eta R_h} \nabla c + \frac{2 R_p^3 \Delta \chi_V}{9 \mu_0 \eta R_h} c \left( \mathbf{B} \cdot \nabla \right) \mathbf{B} + \frac{2 R_p^3 \Delta \rho}{9\eta R_h} c \mathbf{g} + c \mathbf{v}_f
\label{eq:total-flux}
\end{equation}
Therefore, the total flux is a sum of contributions from each of four characteristic forces. To obtain the spatially resolved concentration over time, Eq.~\ref{eq:total-flux} is used with the mass conservation equation: 
\begin{equation}
\nabla \cdot \mathbf{N} = -\frac{\partial c}{\partial t}
\label{eq:mass-conservation}
\end{equation}

The relative effects of the four forces in Eq.~\ref{eq:total-flux} can be compared by up to three independent dimensionless numbers
. First is the magnetic \Peclet number, which compares diffusion with magnetophoresis~\cite{Rassolov2024magnetophoresis}. It is given by the following:
\begin{equation}
\mathrm{Pe}_m = \frac{v_m L}{D} = \frac{4 \pi R_p^3 \Delta \chi_V L}{3 \mu_0 k_B T} \left( \mathbf{B} \cdot \nabla \right) \mathbf{B},
\label{eq:mag-Peclet}
\end{equation}
where $L$ is a relevant length scale. {Prior literature reports the width of the cuvette as a relevant length scale in the dimensionless analysis of induced fluid flow~\cite{Leong2020Langmuir}; therefore, we define $L$ as the width of the cuvette (10~mm) to be consistent with the length scale in our analysis of fluid flow (section \ref{ssc:theory-fluid-flow} below).}

Second is the gravitational \Peclet number, which compares sedimentation with diffusion:
\begin{equation}
\mathrm{Pe}_g = \frac{4/3 \pi R_p^3 \Delta \rho g L}{k_B T}
\label{eq:sed-Peclet}
\end{equation}
{It is primarily relevant when comparing the effects of magnetophoresis with sedimentation; we will return to this point in Sec.~\ref{ssc:magnetophoresis-vs-sedimentation}.

Finally, there is the hydrodynamic (i.e. conventional) \Peclet number, which compares diffusion with induced fluid motion:
\begin{equation}
\mathrm{Pe} = \frac{v_f L}{D}
\end{equation}
}

\subsection{Fluid flow}
\label{ssc:theory-fluid-flow}
Previous studies have demonstrated that magnetophoresis gives rise to concentration gradients that may induce convection~\cite{Leong2015SoftMatter}. For an incompressible Newtonian fluid subjected to a magnetic field gradient, the convection is given by the continuity ($\nabla \cdot \mathbf{v}_f = 0$) and Navier-Stokes equations:
\begin{widetext}
\begin{equation}
\rho \left( \frac{\partial \mathbf{v}_f}{\partial t} + \mathbf{v}_f \cdot \nabla \mathbf{v}_f \right) = (\rho - \rho_l) \mathbf{g} - \nabla \mathcal{P} + \eta \nabla^2 \mathbf{v}_f + \frac{\chi_{V,f}}{\mu_0} (\mathbf{B} \cdot \nabla) \mathbf{B}, \label{eq:Navier-Stokes-magnetic}
\end{equation}
\end{widetext}
where $\rho$ is the density of the suspension, $\rho_l$ is the density of the suspending liquid medium, $\mathcal{P}$ is the dynamic pressure, and $\chi_{V,f}$ is the magnetic susceptibility of the suspension. The last term on the right-hand side of Eq.~\ref{eq:Navier-Stokes-magnetic} is the body force due to the magnetic field gradient. There are five relevant forces: gravitational forces, inertia, viscous forces, forces due to pressure gradients, and magnetic gradient forces. Four independent dimensionless numbers can be obtained. For this system, all are below unit magnitude except for the magnetic Grashof number. The magnetic Grashof number compares the magnetic gradient forces to viscous forces, and is defined by~\cite{Leong2015SoftMatter} as:
\begin{equation}
\mathrm{Gr}_m = \frac{\rho \nabla B \left( \frac{\partial M}{\partial c} \right)_H \left( c_s - c_\infty \right) L^3}{\eta^2},
\label{eqn:Grm-definition}
\end{equation}
where $M$ is the magnetization, $c_s$ is the concentration of the suspended material at the surface, and $c_\infty$ is the concentration in the bulk. For suspensions of paramagnetic and diamagnetic materials, $\mathbf{M}$ is related to $c$ as:
\begin{equation}
\mathbf{M} = \chi_{V,f} \mathbf{H} = \sum_i{\frac{\chi_{m_i}}{\mathcal{M}_i} c_i \mathbf{H}},
\end{equation}
where $\chi_{V,f}$ is the volumetric magnetic susceptibility of the suspension, $\chi_{m,i}$ is the molar magnetic susceptibility of component $i$, $\mathcal{M}_i$ is the molar mass of component $i$, and $c_i$ is the concentration of component $i$. {For the purpose of computing $\chi_V$, the particles and the surrounding water are the two materials that contribute to the magnetization of a fluid element.} The partial derivative in Eq.~\ref{eqn:Grm-definition} evaluates to:
\begin{equation}
\left( \frac{\partial M}{\partial c_i} \right)_H = \frac{\Delta \chi_{m,i}}{\mathcal{M}_i} H = \frac{\Delta \chi_{m,i}}{\mathcal{M}_i \left( 1+\chi_{V,f} \right) \mu_0} B,
\end{equation}
where $\Delta \chi_{m,i}$ is the molar magnetic susceptibility of fluid component $i$ corrected for displacement of the surrounding medium~\cite{note-deltachimol}. 
Therefore, $\mathrm{Gr}_m$ for a suspension of paramagnetic or diamagnetic particles {consisting of one particle material} is:
\begin{equation}
\mathrm{Gr}_m = \frac{\rho \Delta \chi_m \left( c_s - c_\infty \right) L^3}{\mathcal{M}_i (1 + \chi_{V,f}) \mu_0 \eta^2} |(\mathbf{B} \cdot \nabla) \mathbf{B}|_\mathrm{max}
\end{equation}

\section{Materials and methods}

\subsection{Simulations}
\begin{figure}
\centering \includegraphics{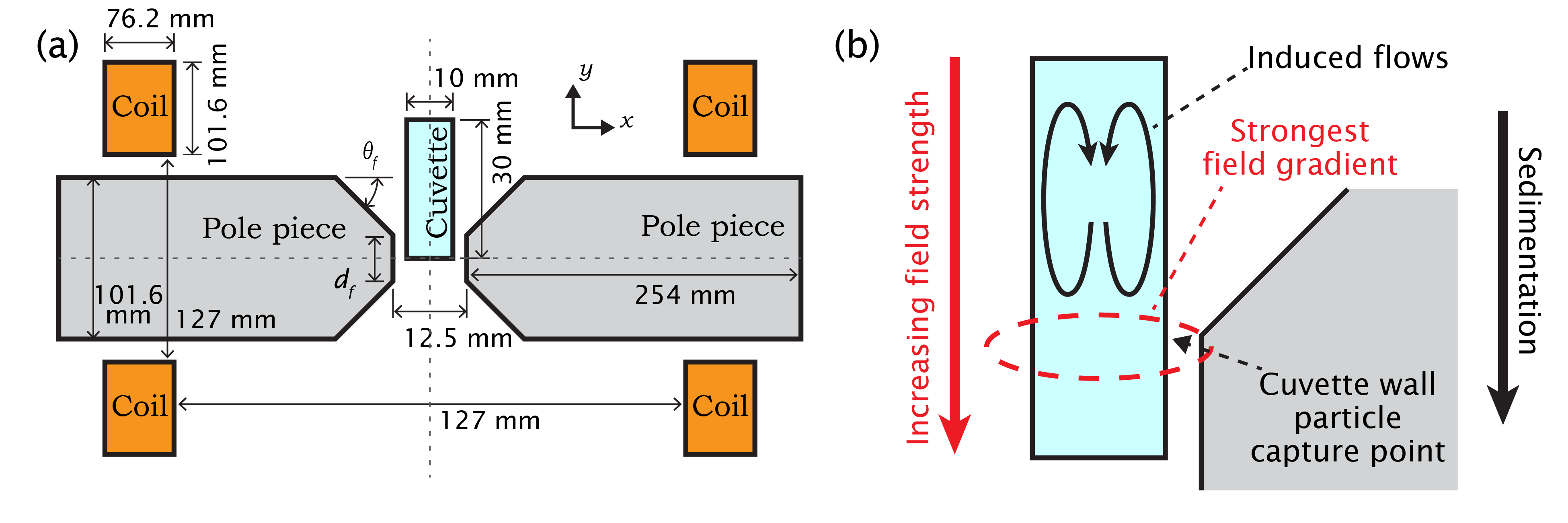}
\caption{\modified{(a)} Schematic of the simulation domain (not drawn to scale). The parameters labeled $\theta_f$ and $d_f$ are varied in the optimization study shown in Fig.~\ref{fig:magnetic-field}(a,b), and the optimized values are shown in this schematic. {(b) Sketch outlining the relevant transport phenomena in this system.}}
\label{fig:simulation-schematic}
\end{figure}

Simulations were completed using COMSOL Multiphysics 6.1. The cuvette and coil apparatus used in the experiments were modeled as a 2-D cross section shown in Fig.~\ref{fig:simulation-schematic}(a). {In addition,  Fig.~\ref{fig:simulation-schematic}(b) shows a schematic of the main forces and transport processes in the cuvette.} The model for cuvette and the coil apparatus was placed at the center of a circular domain of radius 1.5 m. To avoid introducing artifacts related to mesh orientation, the domain representing the cuvette was divided into 0.1~mm $\times$ 0.1~mm square mesh elements. The remaining domains were filled with triangular mesh elements starting around 0.1 mm in size near the cuvette and growing progressively larger with increasing distance from the cuvette. The maximum element sizes were 10 mm in and near the pole piece and coil sections and 200 mm in the air domain. Finally, the circular domain was surrounded by 8 layers of infinite element domain. In total, there were 47,186 mesh elements, of which 30,000 were the square elements inside the cuvette. With an extremely fine mesh size being necessary for accurate 2D simulations, analogous 3D simulations are computationally intractable.

Each simulation consisted of two steps. For the first step, Eqs.~\ref{eq:Amperes-law}--\ref{eq:relative-permeability} were implemented in the Magnetic Fields interface and solved in the steady state to obtain the magnetic field. For the second step, the particle concentration and the fluid velocity field interact; therefore, the equations representing them needed to be solved as a coupled system. For solving the fluid flow equations, no-slip boundary conditions were used on the sides and bottom of the cuvette, and a free surface was used for the top. To represent the capture of particles at the walls of the cuvette due to magnetophoresis and sedimentation, the flux at the boundaries $N_b$ was set as follows:
\begin{equation}
N_b = \max{\left\{ \left[ \frac{2 R_p^3 \Delta \chi_V}{9 \mu_0 \eta R_h} c \left( \mathbf{B} \cdot \nabla \right) \mathbf{B} + \frac{2 R_p^3 \Delta \rho}{9\eta R_h} c \mathbf{g} \right] \cdot \mathbf{n},0\right\} }
\end{equation}
where $\mathbf{n}$ is the outward normal vector at the boundary. Both sets of equations were solved in time-dependent studies with time steps of 0.01 h for the first 3 hours and 0.1 h for the remainder of each simulation (between 4 and 15 h, matching the equivalent experiment duration). 

\subsection{Experiments}
Manganese (III) oxide nanoparticles ($R_p$~=~50~nm) were obtained from US Research Nanomaterials, Inc. and used as received. Dispersions were prepared with 1 g/l nanoparticles in deionized water and stabilized with 10 g/l polyethylene glycol (MW = 6000, Millipore Sigma). To disperse aggregates of particles to the maximum possible extent, the dispersions were sonicated for two hours following preparation. Immediately prior to experiments, the dispersions were again sonicated for 1 hour, then diluted to the desired concentration for experiments in standard disposable cuvettes with 10 mm path length (1 cm $\times$ 1 cm $\times$ 4.5 cm). For each experiment trial, a cuvette containing 3 ml of diluted dispersion was placed in a custom-made holder in the gap between two poles of an electromagnet. The poles and holder were designed to produce the maximum possible \BgradB~in the cuvette; details of the optimization are given in section~(\ref{ssc:magnetic-field-optimization}). The field was measured using a Hall effect magnetometer (Metrolab THM1176-MF) along a radial scan line in the center of the pole gap. Cuvettes were capped to prevent evaporation of the water from the dispersion. The holder was designed such that the center of the bottom of the cuvette was exactly at the axis of symmetry of the electromagnet, where the field was strongest. A light source with a diffusing filter was used to apply a near-uniform incident background light, and a camera was used to acquire a time-lapse sequence of images of the cuvette during each experiment trial. The light intensity was normalized using an initial image of a cuvette containing only deionized water.

\section{Results and discussion}

\subsection{Optimization of magnetic field gradient}
\label{ssc:magnetic-field-optimization}

\begin{figure*}[htb]
\centering \includegraphics[width=1\textwidth]{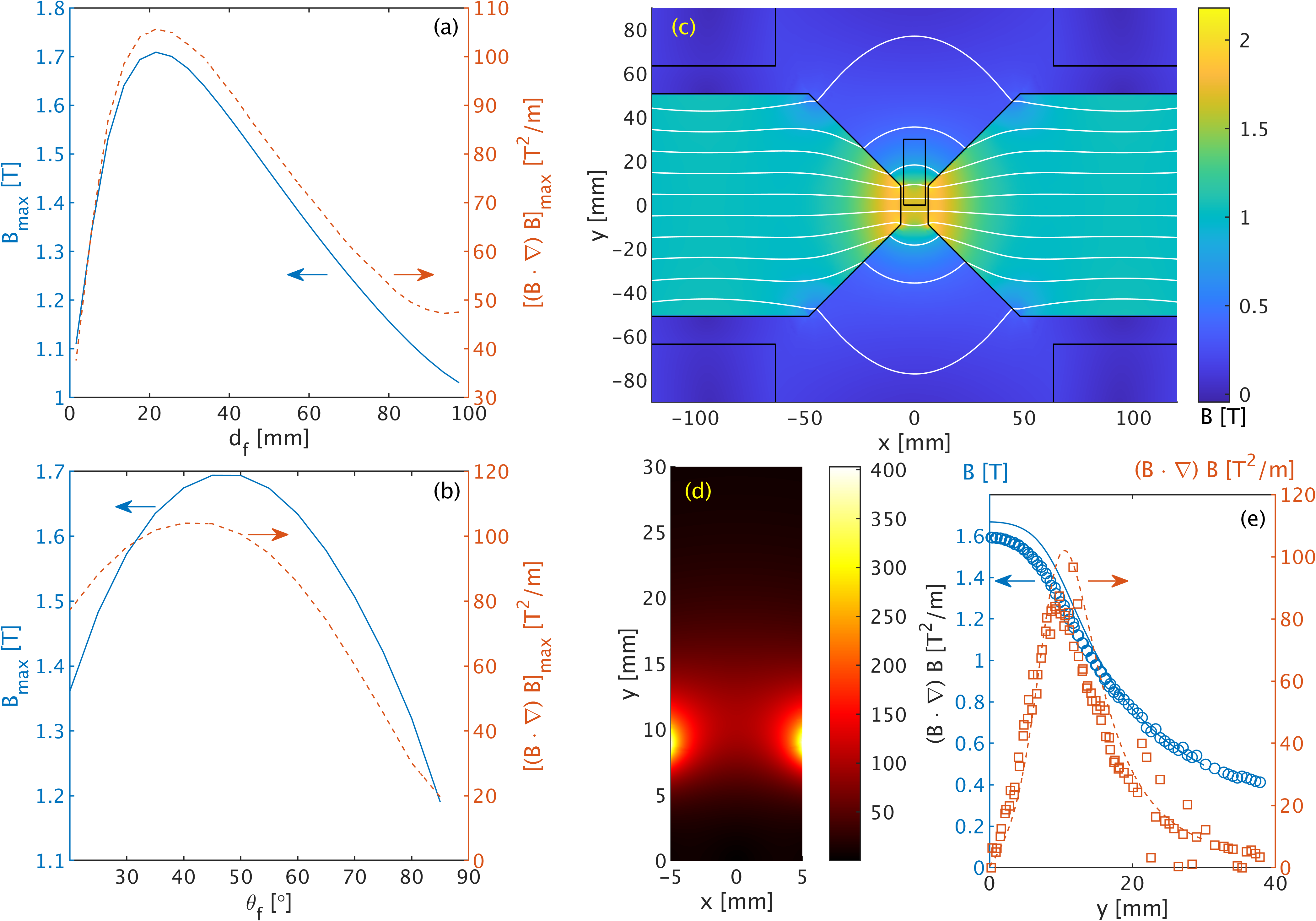}
\caption{Details of the magnetic field in the experiments and simulations. (a) Maximum magnetic field (solid blue line) and field gradient (dashed orange line) along the vertical centerline of the cuvette over different values of pole face diameter $d_f$ with cone angle $\theta_f$ held fixed at 45$^\circ$. (b) Maximum field and gradient over different values of $\theta_f$. (c) Magnitude of $\mathbf{B}$ in and around the pole gap for $d_f$ = 20~mm and $\theta_f$ = 45$^\circ$. The black outlines show the location of the cuvette, pole pieces, and coils, and the white lines are magnetic field lines. (d) Magnitude of $\left( \mathbf{B} \cdot \nabla \right) \mathbf{B}$ inside the cuvette. (e) Simulated $\mathbf{B}$ along the vertical centerline of the cuvette (solid blue line), experimentally measured $\mathbf{B}$ (blue circles), simulated $\left( \mathbf{B} \cdot \nabla \right) \mathbf{B}$ (dashed orange line), and experimentally determined $\left( \mathbf{B} \cdot \nabla \right) \mathbf{B}$ (orange squares).}
\label{fig:magnetic-field}
\end{figure*}
The magnetic field between the two flat pole pieces of the electromagnet is uniform, resulting in negligible magnetophoresis. To enhance magnetophoretic effects, we focused on increasing the magnetic field gradients achievable with the 1T electromagnet. This optimization was carried out using 2D simulations, where the magnetic field was calculated as a function of the pole size, 
$d_f$ and the pole angle $\theta_f$. Figure~\ref{fig:magnetic-field} (a) presents the variation of $\mathbf{B}$ (primary y-axis), and $\left( \mathbf{B} \cdot \nabla \right) \mathbf{B}$ (secondary y-axis) as a function of $d_f$ for a vertical line that is aligned with $x = 0$; the middle of the cuvette. Based on these simulations, $d_f \approx$  20 mm produces the maximum magnetic field gradients.
Figure~\ref{fig:magnetic-field}(b) shows the magnetic flux density and the magnetic field gradients as a function of $\theta_f$. These results suggest that $\theta_f \approx $  45$^\circ$ produces the maximum magnetic field gradients. Figure~\ref{fig:magnetic-field}(c,d) present the simulated magnetic flux density and $\left( \mathbf{B} \cdot \nabla \right) \mathbf{B}$ inside the cuvette for the most optimal pole dimensions. The conical pole pieces concentrate magnetic flux into the gap between the truncated ends, resulting in a magnetic flux density of 1.6 T. This strong and localized field gives rise to $\left( \mathbf{B} \cdot \nabla \right) \mathbf{B}$ as high as 380~T$^2$/m. Finally, Figure~\ref{fig:magnetic-field}(e) shows that the comparison between the simulated magnetic field for the optimal pole dimensions and the measured values for the same pole dimensions obtained from experiments. The simulation results are within 5\% of the experimentally measured field, demonstrating a very good agreement with the experimental data. The slight overprediction of the magnetic flux density at the field center by simulations is presumably due to the approximation of the apparatus as a planar cross-section.\par 

{\subsection{Particle transport under no magnetic field}
Before conducting magnetophoresis experiments and simulations, we performed experiments without a magnetic field for two main reasons: first, to assess the impact of particle sedimentation and second, to validate our model against experimental results obtained in the absence of a magnetic field. Using the particle size and density provided by the vendor, we can evaluate the gravitational Peclet number as $Pe_g \approx 2.5\times10^{-4}$, suggesting that sedimentation of particles should not be significant.\par 
\begin{figure}[ht]
\centering \includegraphics{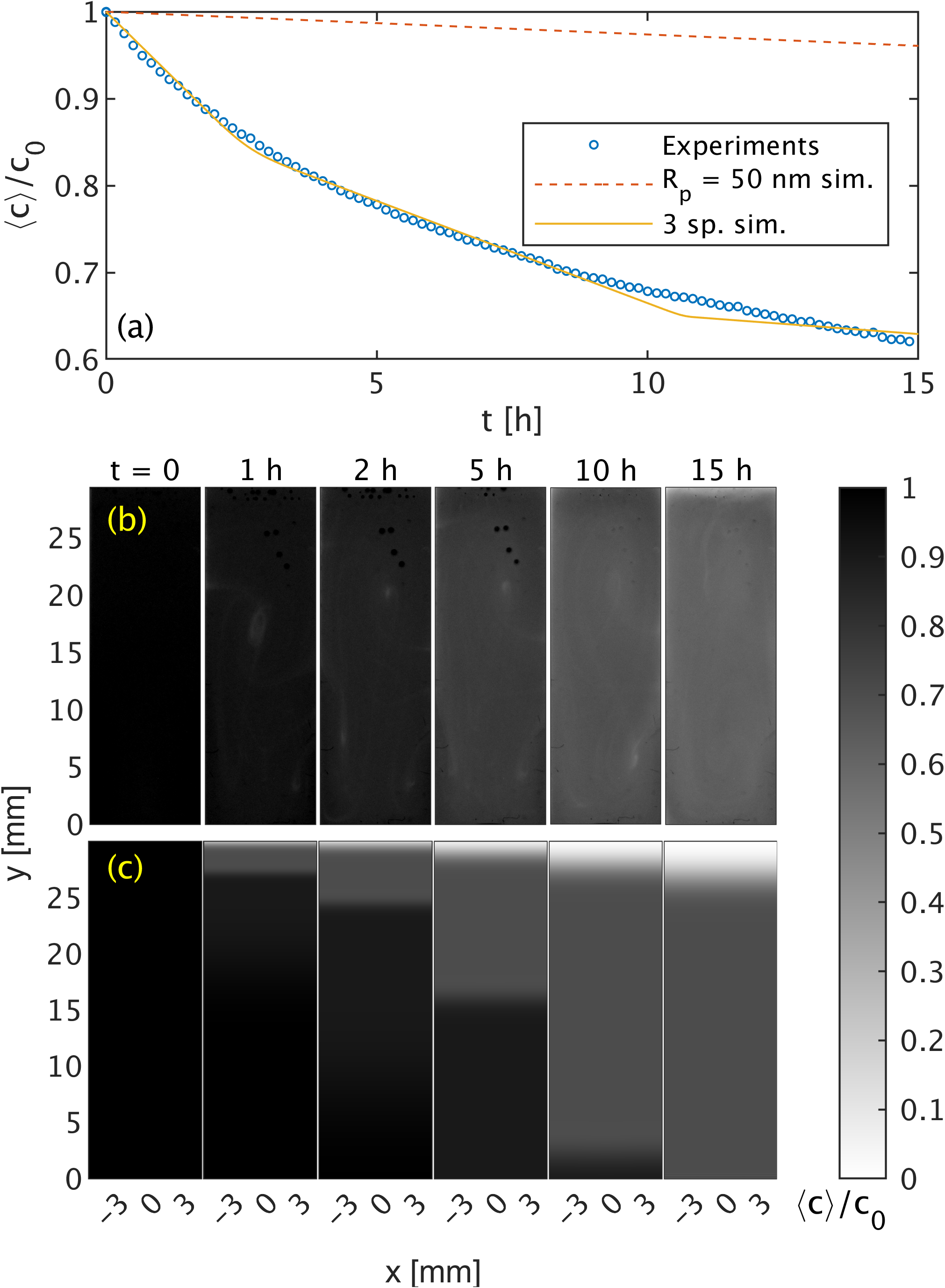}
\caption{Temporal evolution of particle sedimentation from the cuvette with no applied field and $c_0$~=~100~mg/l. (a) Average normalized concentration of particles observed in experiments (blue circles), predicted in simulations for monodisperse particles as specified (dashed orange line{; identified as 50 nm sim.}), and predicted for the size distribution in Table~\ref{tbl:sim-size-dist} over the duration of the experiment (solid yellow line{; identified as 3 sp. sim.}). (b) Experimentally determined spatially resolved concentration at selected times. (c) Simulated spatially resolved concentration at the same times as in (b) for the particle size distribution in Table~\ref{tbl:sim-size-dist}.}
\label{fig:evol-field-off}
\end{figure}

Fig.~\ref{fig:evol-field-off} (a) shows the temporal evolution of the normalized spatially averaged concentration of manganese oxide particles in a cuvette without an applied magnetic field in both experiments and simulations. The concentration of particles is calculated as:
\begin{equation}
c = \frac{A}{\langle A_0 \rangle} c_0,
\end{equation}
where $A$ is the light absorbance of the solution at any pixel location (see SI for calculation of the absorbance and validation of the Beer-Lambert law for this system) and $\langle A_0 \rangle$ is the average light absorbance over all pixels containing the solution for the initial image. The spatially averaged concentration $\langle c \rangle$ is simply the average of the calculated $c$ throughout all pixel locations in the image captured at time $t$. Figure~\ref{fig:evol-field-off}(a) shows that experimentally determined average concentration decreases slowly over time through sedimentation, with 62\% of the particles remaining suspended in the fluid after 15 hours. Interestingly, simulations with a single particle size of $R_p =$ 50 nm (represented by the dashed line) underpredict the experimental results. We hypothesize that these particles may have formed aggregates, leading to faster sedimentation. Additionally, if the particle size distribution in the solution is narrow, the temporal evolution of concentration should follow first-order kinetics, resulting in a straight line on a semi-logarithmic scale. However, the experiments show that the sedimentation rate is initially fast and then gradually slows down. We hypothesize that this behavior is due to formation of particle aggregates with a broader size distribution. To test these hypotheses, we performed numerical simulations using a particle size distribution that follows $\sum_{i=1}^{3} f_i R_{pi}$, where {$R_{pi}$ is the radius of a particle of one size class and $f_i$ is the mass fraction of that size class relative to all particles.} Note that in the simulations, the particle size distribution was modeled as a 3 species mixture, as simulations with a broader particle size distribution become computationally intractable. The continuous line in Fig.~\ref{fig:evol-field-off}(a) shows the best match between experiments and simulations. The resulting particle size and their mass fractions are reported in Table~(\ref{tbl:sim-size-dist}).}{ To independently assess the particle size distribution under no magnetic field, we conducted dynamic light scattering (DLS) measurements. While DLS is inherently more sensitive to larger particles due to its intensity-weighted nature, the measured hydrodynamic radius was $R_p = 330 \pm 30$ nm with a polydispersity index of approximately 0.3. The particle size distribution inferred from simulations, which features a dominant size class around 80 nm and a long tail extending up to 600 nm, aligns with the DLS results in indicating the presence of a significant population of larger particles.} The sequence of images in Figure~\ref{fig:evol-field-off}(b) shows the spatially resolved experimentally determined concentration at selected times. A slight concentration gradient from lowest at the top to highest at the bottom develops after approximately 10 h. Figure~\ref{fig:evol-field-off}(c) shows the simulation results; the apparent sharp horizontal interfaces arise from the discretization of the particle size distribution into 3 species.\par 
\subsection{Magnetophoresis} \label{ssc:magnetophoresis-results}
\begin{figure}[ht]
\centering \includegraphics{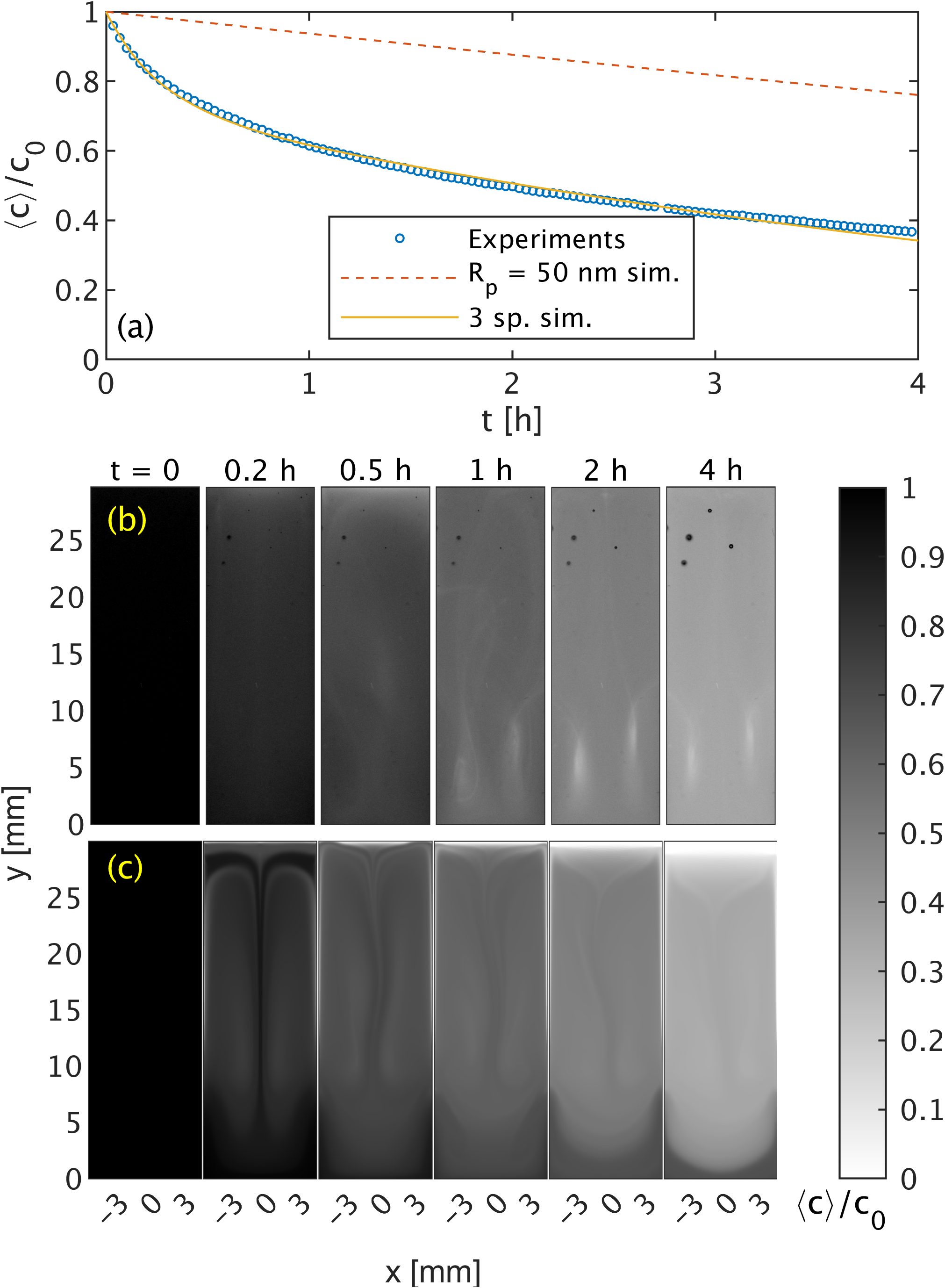}
\caption{Temporal evolution of particle removal from the cuvette with the full applied magnetic field and $c_0$ = 100mg/l. (a) Average normalized concentration of particles observed in experiments (blue circles), predicted in simulations for monodisperse particles as specified (dashed orange line), and predicted for the size distribution in Table~\ref{tbl:sim-size-dist} over the duration of the experiment. The light gray markers and line are the experimentally observed and 3-species simulation results shown in Fig.~\ref{fig:evol-field-off} and are included in this plot for comparison. (b) Experimentally determined spatially resolved concentration at selected times. (c) Simulated spatially resolved concentration at the same times as in (b).}
\label{fig:evol-field-down}
\end{figure}

Figure~\ref{fig:evol-field-down}(a) shows the temporal evolution of the manganese oxide particle concentration at a fixed initial concentration of 100~mg/l and at full applied magnetic field. The rate of particle removal is greatly increased by applying the magnetic field; only 37\% of particles remain suspended after 4 h. For comparison under no magnetic field 80\% of particles are still suspended in the domain (see -o data in Fig.~\ref{fig:evol-field-off}(a)). {Figure~\ref{fig:evol-field-down}(b) presents a sequence of experimental snapshots showing the evolution of normalized particle concentration over time (see Video 1 in the supplementary materials for the full 4-hour experiment). As time progresses, a magnetophoretically induced flow emerges, eventually permeating the entire cuvette. This flow often gives rise to multiple vortical structures characterized by locally depleted particle concentrations dispersed throughout the domain. Notably, after approximately one hour, two distinct low-concentration regions develop around $y=5$~mm and $x=\pm 3$~mm. These locations coincide with the zones of maximum magnetic field gradient \BgradB, as indicated in Fig.~\ref{fig:magnetic-field}(d). The magnetic field gradient changes most rapidly near the point where it is highest, and this rapid change results in rapid, localized depletion of particles as they are removed from the suspension. In experiments, they adhere to the walls of the cuvette. The migration of particles from areas of lower magnetic field is expected to occur more slowly. Note that the results of these experiments are different than those of superparamagnetic nanoparticles where a uniform concentration profile was reported in the cuvette~\cite{Leong2015SoftMatter}. We will return to the nature of such differences in Section~(IV D) below. } \par 

Similarly to what was found under no magnetic field, simulations with assumed monodisperse 50 nm particles {(shown as the dashed line Fig.~\ref{fig:evol-field-down}(a))} underpredict the observed particle removal by a wide margin, and a best matching 3-species distribution was identified. The simulation results for the best 3-species distribution is shown as a continuous curve in Fig.~\ref{fig:evol-field-down}(a) and Fig.~\ref{fig:evol-field-down}(c). {The particle size distribution that yields the best agreement with experimental observations was determined through a trial-and-error approach and is summarized in Table~\ref{tbl:sim-size-dist}. The distribution is notably skewed toward larger particle sizes, which may indicate the formation of field-induced clusters. This interpretation and its implications are addressed in detail later in the manuscript.} {In addition, Figure~\ref{fig:evol-field-down}(c) illustrates the spatio-temporal evolution of normalized particle concentration from simulations using the particle size distribution that best matches the experimental data shown in Fig.~\ref{fig:evol-field-down}(a). During the transient phase, the simulations predict the formation of induced convective flows spanning the full height of the cell. These flows lead to localized regions of reduced particle concentration, particularly in the range $y = 10\text{–}15$ mm and $x = \pm 2$ mm, emerging as early as t = 0.2 h. As time progresses, these convective patterns facilitate gradual particle separation and capture by the wall. It is worth noting that the low-concentration regions predicted by the simulations appear earlier and are less pronounced compared to those observed experimentally. Additionally, the simulated concentration gradients, characterized by depleted regions near the top and accumulation near the bottom corners are not observed experimentally. These differences are likely influenced by the simplification of modeling the cuvette as a two-dimensional rectangular domain, and possibly other reasons that are discussed toward the end of the manuscript. }\par

\begin{table}
\centering
\begin{tabular}{|c c|c c|}
\hline
\multicolumn{2}{|c|}{\BgradB~= 0 T$^2$/m} & \multicolumn{2}{c|}{\BgradB~= 110 T$^2$/m} \\ \hline
$R_p$ [nm] & $f$ & $R_p$ [nm] & $f$ \\ \hline
80 & 0.76 & 80 & 0.64 \\
300 & 0.16 & 300 & 0.24 \\
600 & 0.08 & 800 & 0.12 \\ \hline
\end{tabular}
\caption{Particle size distributions for simulations in Figs.~\ref{fig:evol-field-off} and \ref{fig:evol-field-down}. {$f$ is the fraction of particles (by mass) of each size class.}}
\label{tbl:sim-size-dist}
\end{table}

\begin{figure}[ht]
\centering \includegraphics{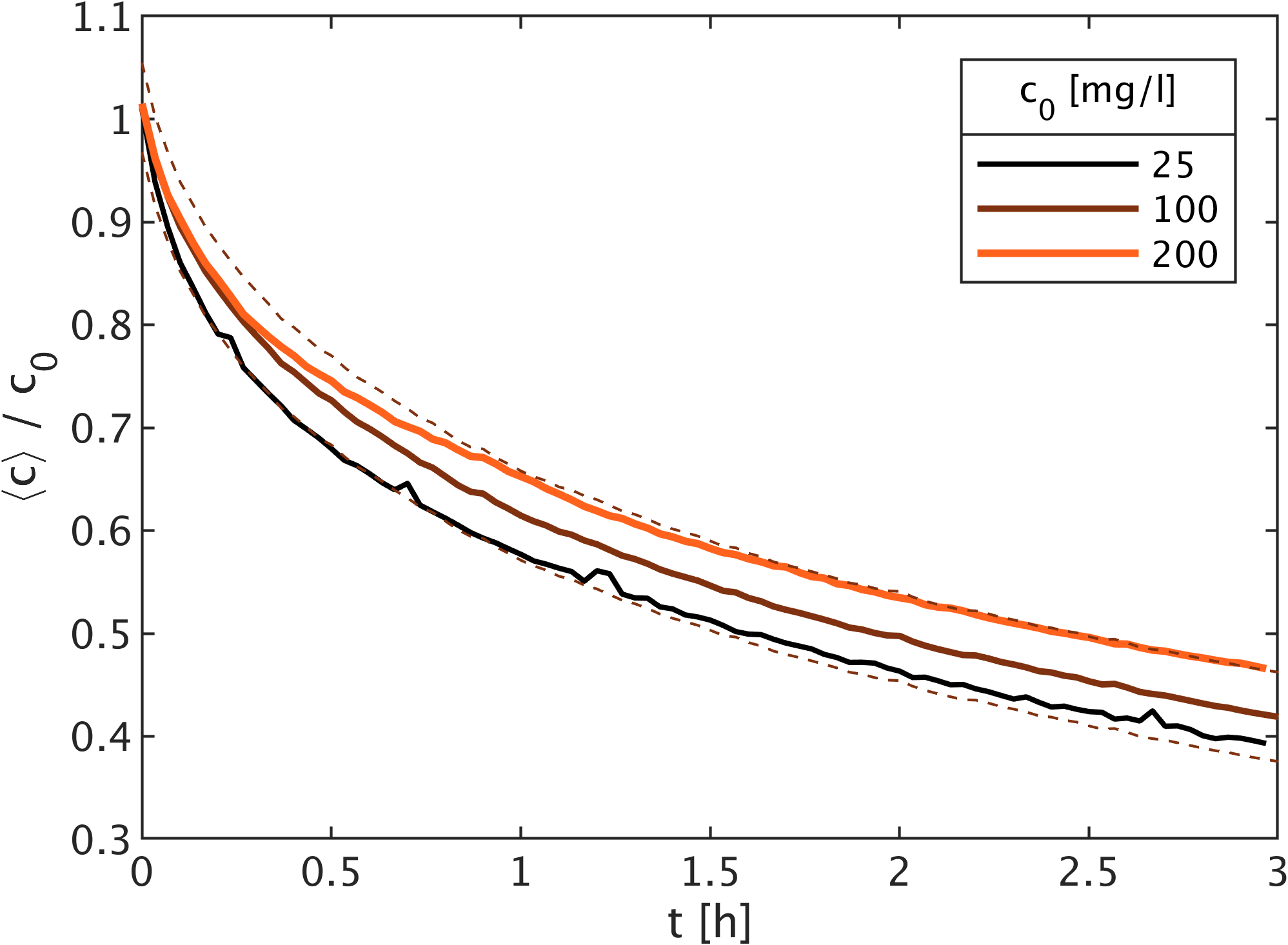}
\caption{Temporal evolution of the spatially averaged normalized concentration of manganese oxide nanoparticles for varied initial concentration with the full applied magnetic field. {The dashed lines indicate the standard deviation from replicate experiment trials with $c_0 = 100 \mathrm{mg/l}$.}}
\label{fig:effect-of-c0}
\end{figure}

Figure~\ref{fig:effect-of-c0} shows the temporal evolution of the manganese oxide particle concentration at various initial concentrations and at full magnetic field. {For the range of concentrations studied, the differences in average normalized concentration evolution over time between different values of initial concentration are within experimental uncertainty.} These results are consistent with those results of Leong et al.~\cite{Leong2015SoftMatter} on superparamagnetic nanoparticles based on iron oxide. \par 

\begin{figure}[h]
\centering \includegraphics[scale = 0.95]{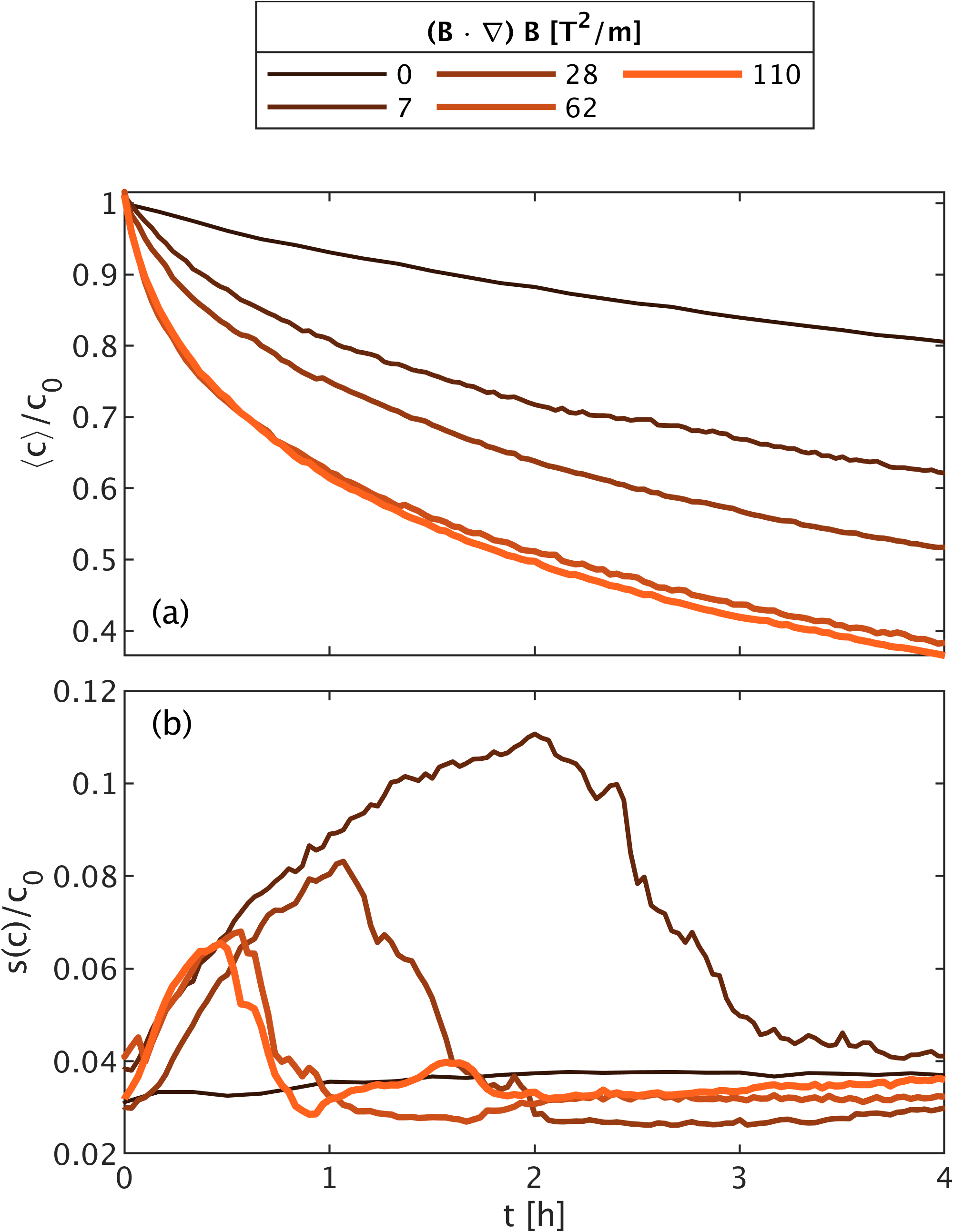}
\includegraphics[scale = 0.95]{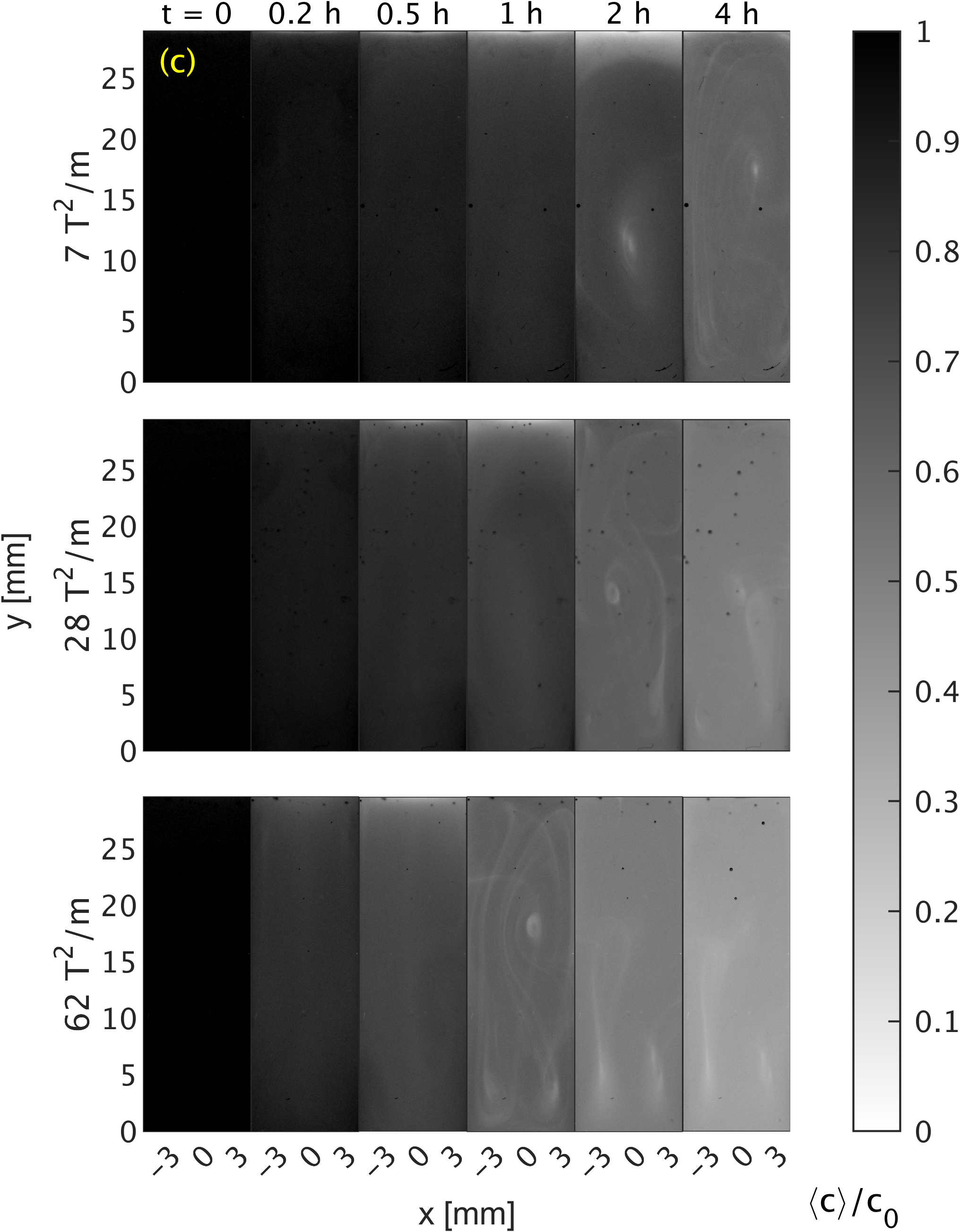}
\caption{(a) Temporal evolution of particle concentration as a function of \BgradB at initical concentration $c_0$ = 100mg/l. {(b) Experimentally determined spatially resolved concentration at selected times under varied magnetic field strength and $c_0 = 100~\mathrm{mg/l}$.} (c) Temporal evolution of the standard deviation of the spatially resolved concentration of manganese oxide particles compared for different \BgradB. The standard deviation is calculated for all pixels in the image captured at time $t$; higher values of $s(c)$ indicate nonuniform distribution of $c$ throughout the cuvette. The values given in the legend are the maximum \BgradB~along the vertical centerline of the cuvette.}
\label{fig:conc-evol}
\end{figure}

To further confirm that the enhanced particle transport is associated with the magnetophoretic force, we systematically varied the external magnetic field (or the associated maximum \BgradB.) Fig.~\ref{fig:conc-evol}(a) shows the temporal evolution of particle concentration measured as a function of \BgradB.{ As the magnetic field gradient increases, the separation dynamics are enhanced; however, at sufficiently high gradients, the variation in particle concentration reaches a plateau. To better understand this behavior, we analyze the spatio-temporal evolution of particle concentration shown in Fig.~\ref{fig:conc-evol}(b). At low gradients (\BgradB~ = 7 T$^2$/m), magnetophoretically induced convection is relatively weak, resulting in the formation of a single vortex accompanied by modest particle depletion near the center of the cuvette after 2 hours, along with a more pronounced particle-depleted region near the top (visible at t = 1 and 2 hours). While this concentration gradient gradually dissipates at later times, a weak convective flow persists throughout the cuvette.\par 

As the field gradient increases to \BgradB~= 28 T$^2$/m, the induced convective flow becomes more pronounced, particularly after 1 hour. At higher gradients \BgradB~ = 62 T$^2$/m, the convection intensifies significantly, leading to more homogenized concentration profiles across the cuvette. After 1 hour, two well-defined vortices emerge near regions of maximum magnetic field gradient. Further increasing the gradient to \BgradB~ = 110 T$^2$/m does not produce qualitatively different convective patterns compared to \BgradB~ = 62 T$^2$/m, suggesting a saturation in the magnetophoretic convective response. This saturation is consistent with the plateau observed in Fig.~\ref{fig:conc-evol}(a), where particle concentration variation no longer increases with field strength. \par 

To further quantify the spatial uniformity in concentration of particles noted in Fig.~\ref{fig:conc-evol}(b), we plot the temporal evolution of standard deviation of the spatially resolved concentration for varied magnetic field strengths.} When no magnetic field is applied, the standard deviation of the concentration rises slightly, from about 3\% of the initial concentration (likely due to measurement noise) to near 4\% as time progresses. With the field applied, the standard deviation initially increases, goes through a maximum, and then decreases to a steady state value near the initial value. The extent of the increase is greatest where \BgradB$|_{max}$~= 7 T$^2$/m, the lowest nonzero value studied, and it occurs after two hours of magnetic field exposure. For stronger fields, the maximum in the standard deviation occurs sooner and becomes less significant, although the overall temporal trend remains consistent. {Taken together, the results in Fig.~\ref{fig:conc-evol}(a–c) reveal a strong coupling between overall concentration depletion in the cuvette, the strength of the induced convective flow, and the resulting concentration gradients developed within the cuvette. Specifically, the observed saturation in concentration dynamics at high magnetic field gradients is likely a consequence of the saturation in the strength of the magnetophoretically induced flow.} In their study of magnetophoresis of superparamagnetic iron oxide nano particles, Leong et al. noted that the distribution of particles remains uniform throughout the bulk fluid with no detectable concentration gradient~\cite{Leong2015SoftMatter}. Our results of Figs.~\ref{fig:conc-evol}(b) show that for the manganese oxide particles studied here, concentration gradients do form. 

\begin{figure}[ht]
\centering \includegraphics{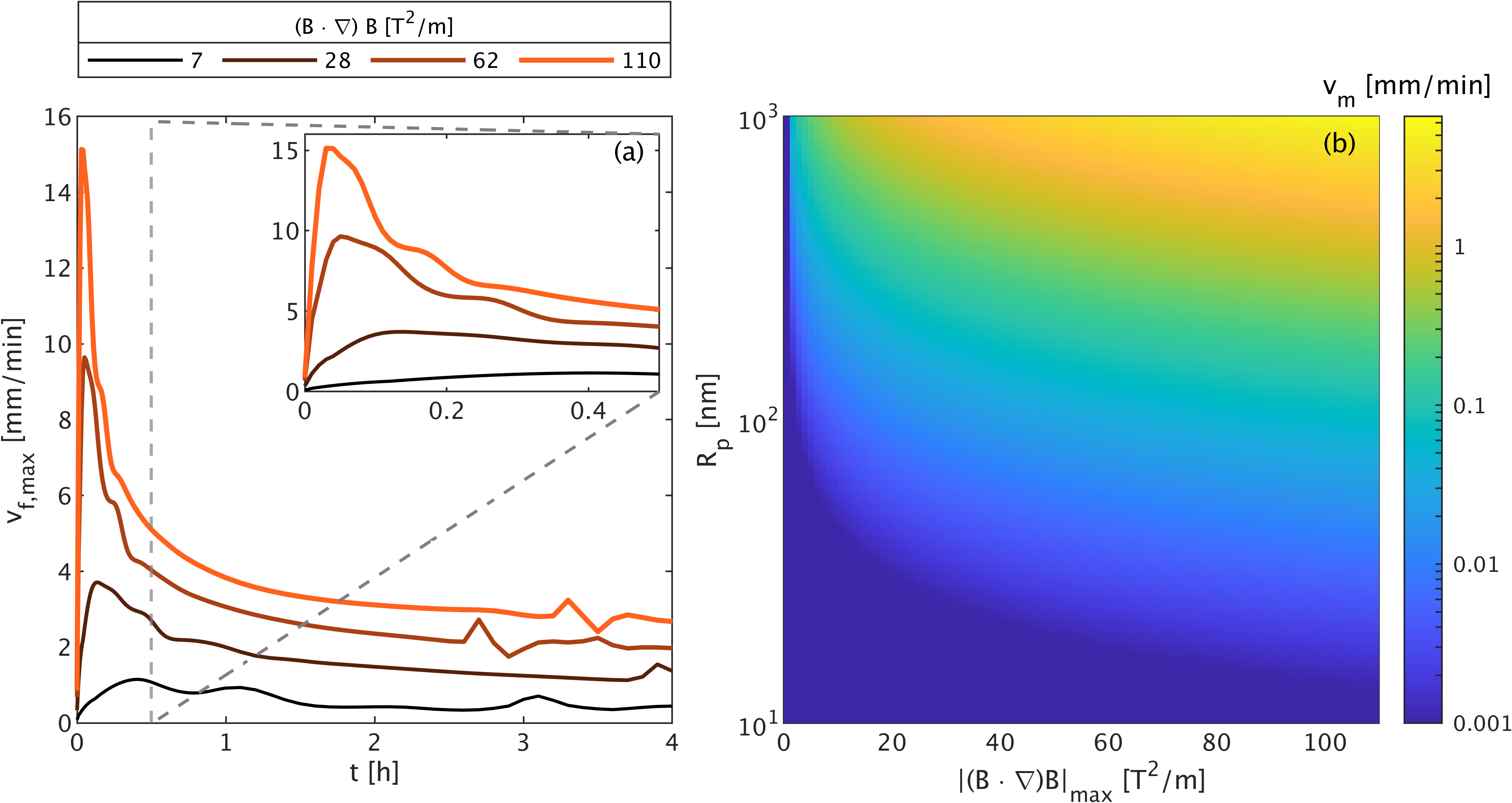}
\caption{{(a) Simulated fluid velocity magnitude at the point of fastest flow over time for varying field strengths. (b) Magnetophoretic velocity as calculated as a function of magnetic field gradients and particle size using Eq.~\ref{eq:magph-velocity}.}}
\label{fig:velocity-evolution}
\end{figure}

{\subsection{Magnetophoretic induced convective flows}}\label{ssc:magnetophoresisconvection}
To gain deeper insight into the origin of the concentration gradients observed in our experiments and the saturation of particle transport at high magnetic field gradients (as shown in Fig.~\ref{fig:conc-evol}), we conducted a detailed analysis using {the results from the simulations reported in Sec.~\ref{ssc:magnetophoresis-results}. Figure~(\ref{fig:flow}) illustrates the spatio-temporal evolution of the velocity field generated by magnetophoretically induced convection within the cuvette under varying magnetic field gradients. At low field strength (\BgradB~=~7 T$^2$/m), two circulation zones (vortices) emerge near regions of maximum magnetic field gradient. The formation of such induced flows is expected, as the inhomogeneous magnetic field applies a spatially varying force on the fluid, leading to faster depletion of nanoparticles in the lower regions of the solution compared to the upper regions. This differential transport generates a vertical concentration gradient. When this stratification occurs in the presence of a vertically decreasing magnetic field gradient, it introduces a mechanical instability in the system. Such instabilities are typically alleviated through the onset of convective motion, as previously reported~\cite{Leong2020Langmuir}. Shortly after the onset of the induced flow, the well-defined vortices begin to transfer momentum to the surrounding fluid, expanding vertically to reach the top of the cuvette. This leads to the formation of an impinging jet that descends along the central axis of the domain. Over time, the strength of the vortices diminishes, but their structure remains discernible throughout the cell. As the magnetic field gradient increases, the strength of the vortices and the associated flow velocities increase markedly, although the overall spatio-temporal evolution retains a similar qualitative pattern. \par 
\begin{figure}[hthp]
\centering \includegraphics{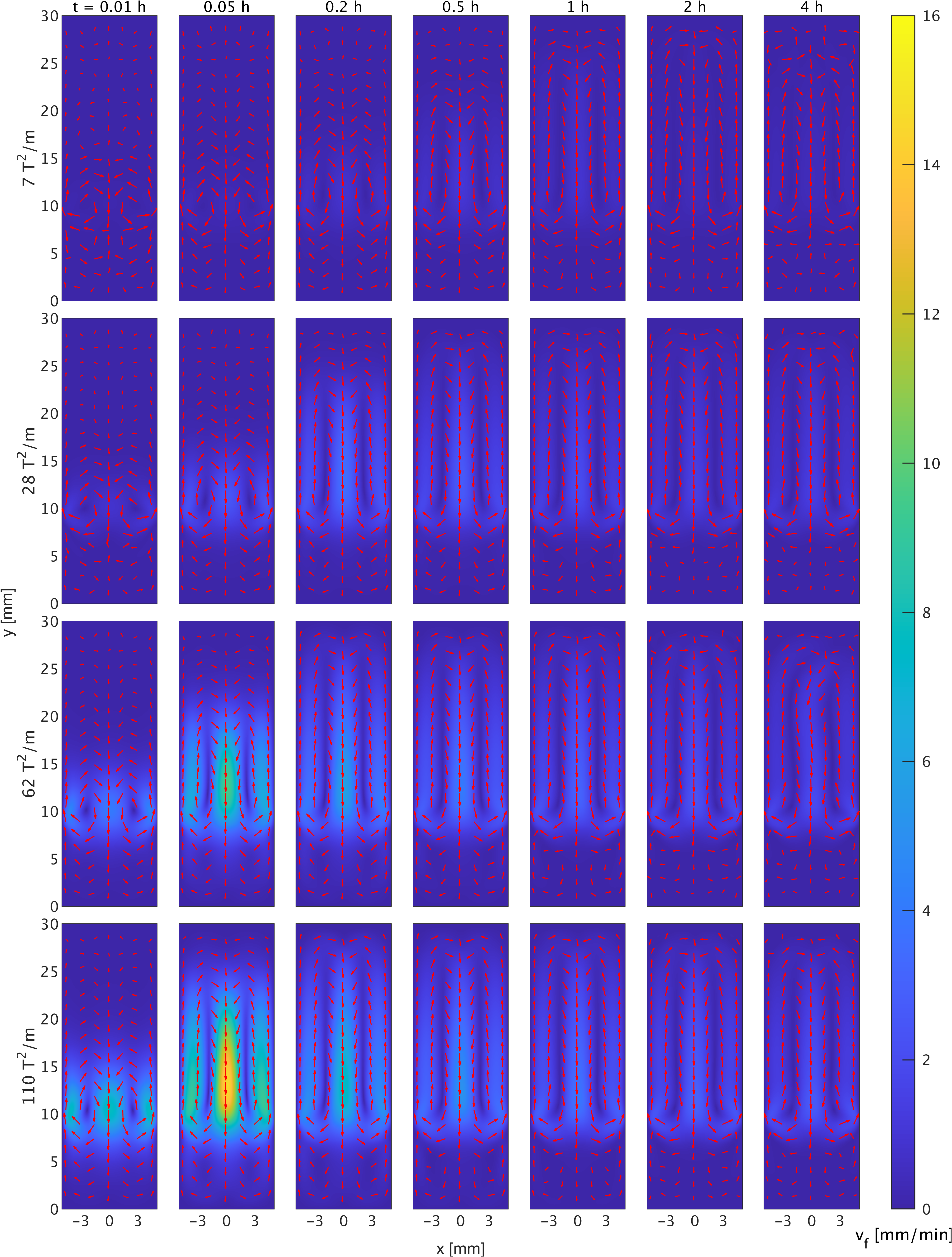}
\caption{{Simulated spatially resolved fluid velocity vectors at selected times for varying field strengths. }}
\label{fig:flow}
\end{figure}

Figure~\ref{fig:velocity-evolution}(a) presents the temporal evolution of the maximum induced fluid velocity ($v_{f,max}$) in simulations of Fig.~\ref{fig:flow} for a range of magnetic field gradients. At the lowest gradient (\BgradB~ = 7 T$^2$/m), the induced flow yields a velocity of 1mm/min, which leads to a Péclet number of $\mathrm{Pe} = 3.8 \times 10^4$, highlighting the dominance of advective transport over diffusion. Notably, even at this low field strength, the induced fluid velocity significantly exceeds the magnetophoretic particle velocity $v_m$ calculated and shown in Fig.~\ref{fig:velocity-evolution}(b). As the magnetic field gradient increases, the maximum induced velocity rises correspondingly, with Péclet numbers reaching up to $\mathrm{Pe} \approx 6 \times 10^5$ at \BgradB~ = 110 T$^2$/m. {A notable feature observed in these simulations is the emergence of a transient maximum in the magnetophoretically induced fluid velocity. This behavior can be interpreted as a consequence of the dynamic evolution of particle concentration within the flow cell. At the initial time point ($t=0$), the particle concentration is spatially uniform and equal to the initial value $c_0$, and thus no concentration gradients exist to drive fluid motion. During a short induction period, particles located near the magnetic capture zone are removed, establishing a concentration gradient that initiates flow via magnetically driven convection. As the flow evolves, particles from the bulk are advected toward the capture region and removed, steepening the concentration gradient and temporarily amplifying the magnetophoretic driving force. This positive feedback leads to an increase in fluid velocity. However, as particle depletion progresses and the overall concentration in the domain declines, the concentration gradient weakens, causing the driving force, and thus the fluid velocity, to decrease. This results in a transient peak in magnetophoretic velocity, followed by a gradual reduction over time.} Although the fluid velocity decreases and eventually asymptote to a smaller value, the fluid velocity at longer times is still significant underscoring the importance of induced convective flow on transport of particles under stronger magnetic gradients.}\par

\begin{figure}[ht]
\centering \includegraphics{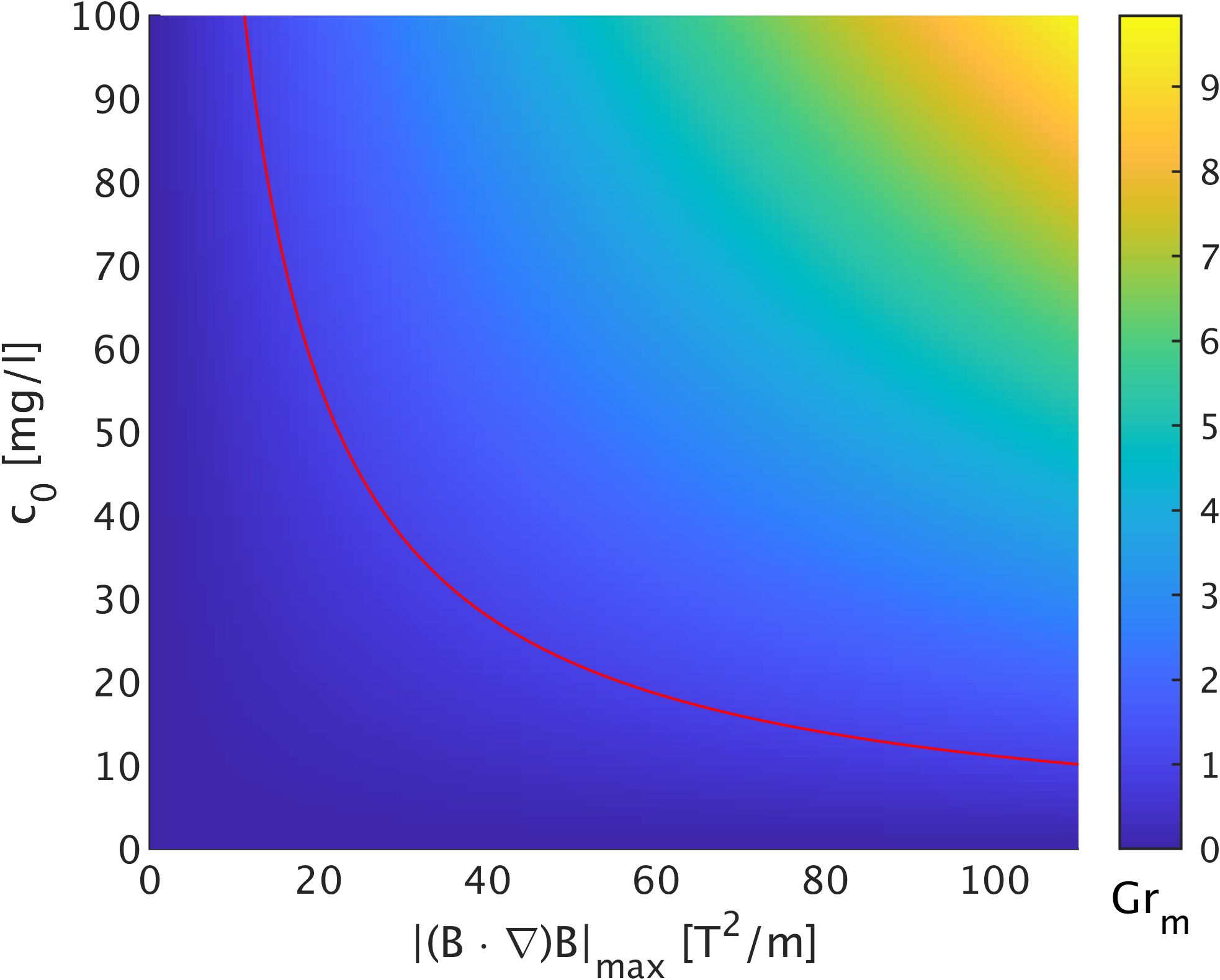}
\caption{Variation of the magnetic Grashof number with \BgradB~and initial concentration. The red line marks $\mathrm{Gr}_m = 1$.}
\label{Grashof}
\end{figure}
{Finally, we connect these induced convective flows with the concentration gradients formed in the cuvette as plotted in Fig.~\ref{fig:conc-evol}. To accomplish this, we examined the variation of the magnetic Grashof number $\mathrm{Gr}_m$ over a broad range of initial particle concentrations and \BgradB. Recall that the dimensionless magnetic Grashof number, $\mathrm{Gr}_m$, quantifies the relative importance of magnetically induced flows compared to viscous flow. These induced flows oppose the development of steep concentration gradients by enhancing mixing within the domain. Consequently, as $\mathrm{Gr}_m$ increases, the ability of the system to sustain strong concentration gradients diminishes, leading to more uniform particle distributions. Fig.~\ref{Grashof} shows the calculated magnetic Grashof number over a broad range of particle initial concentration and \BgradB$|_{max}$.} At very low magnetic field gradients (i.e., \BgradB ~= 7 T$^2$/m), where $\mathrm{Gr}_m = 0.6$, the secondary convective flows are anticipated to be weak and less influential. Under these conditions, a concentration gradient is expected to develop within the cuvette, with particle concentrations being higher in regions exposed to stronger magnetic field gradients. As \BgradB~ increases further beyond 7 T$^2$/m, the intensification of convective flows (corresponding to higher $\mathrm{Gr}_m$ enhances bulk fluid motion, effectively diminishing the concentration gradients at an accelerated rate. 
As \BgradB~continues to increase, the growth rate of $\mathrm{Gr}_m$ decelerates, approaching an asymptotic value of approximately $\mathrm{Gr}_m \approx$ 9 as {\BgradB~approaches 100~T$^2$/m}. Consequently, the spatial variation in particle concentration across the domain is expected to stabilize and level off. The above analysis is consistent with the experimental observations shown in Fig.~\ref{fig:conc-evol}(a,c). {Furthermore, the observed similarities between the curves in Fig.~\ref{fig:conc-evol}(a) and Fig.~\ref{fig:conc-evol}(c) motivated us to apply a scaling analysis to collapse the data in terms of time and the evolving concentration gradients. Notably, these scaling factors exhibit power-law dependencies on the magnetic Grashof number $\mathrm{Gr}_m$. Specifically, the characteristic time required to establish a steady concentration gradient within the cuvette scales as $t\sim \mathrm{Gr}^{-1/2}_{m}$, while the normalized spatial standard deviation of concentration follows $[s(c)-s_0]/c_0\sim \mathrm{Gr}^{-1/4}_{m}$ (see Fig.~S2 in the Supplementary Information). }

Finally, the $\mathrm{Gr}_m$ values reported by Leong et al.~\cite{Leong2015SoftMatter} ($\mathrm{Gr}_m > 100$) are significantly higher than those calculated in this study, which explains the minimal concentration gradients observed in their experiments. To the best of our knowledge, this is the first study where concentration gradients of paramagnetic particles are observed in experiments.\par 

\subsection{Competition between magnetophoresis and sedimentation}
\label{ssc:magnetophoresis-vs-sedimentation}
\begin{figure}[ht]
\centering \includegraphics{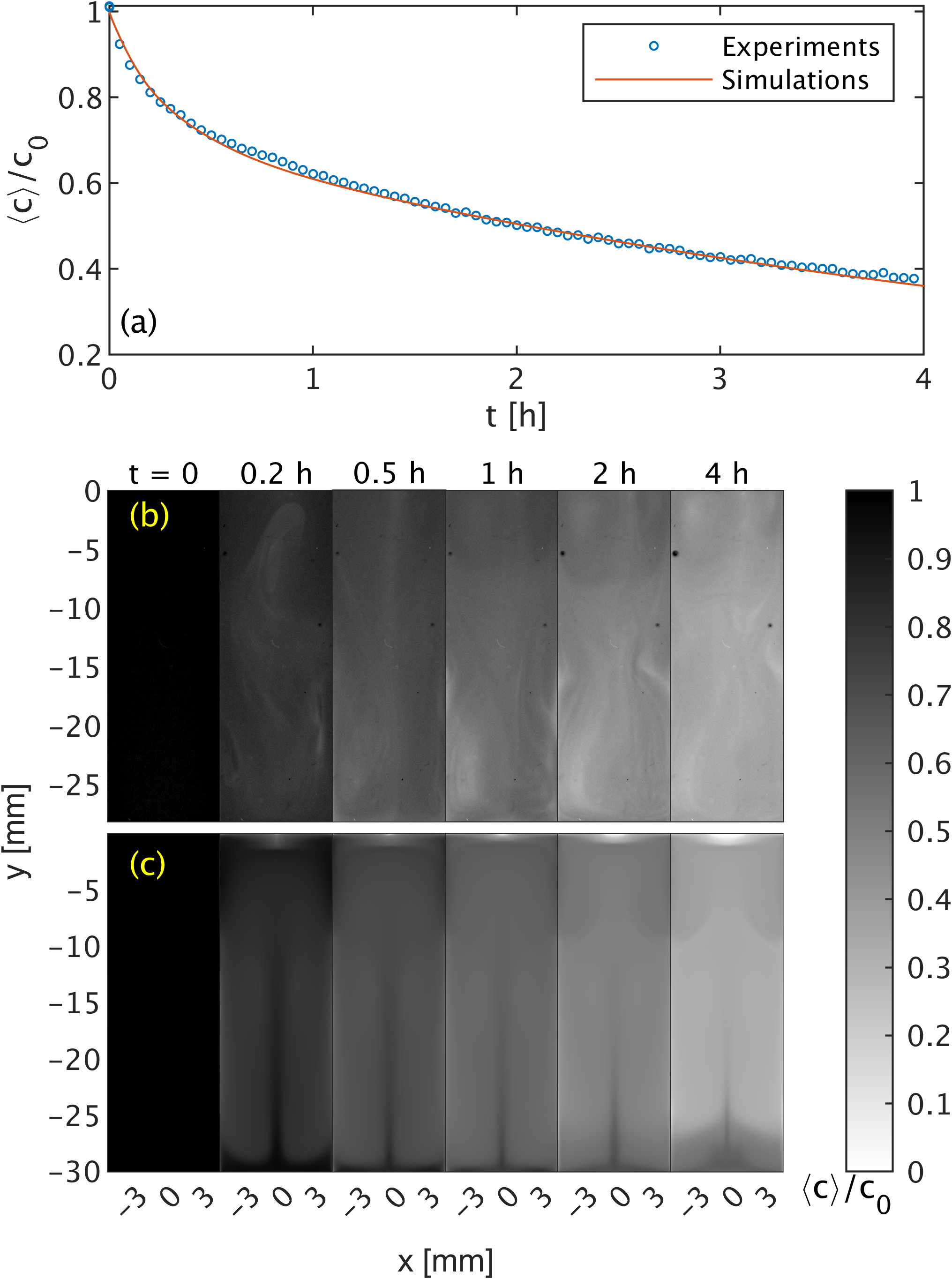}
\caption{Temporal evolution of particle removal from the cuvette with the direction of the magnetic field reversed along the $y$ axis at a fixed initial concentration $c_0$ = 100mg/l and \BgradB ~= 110 T$^2/$m. (a) Average normalized concentration of particles for experiments and simulations over the duration of the experiment. The light gray markers and line are the concentrations for the original direction of the magnetic field shown in Fig.~\ref{fig:evol-field-down}. (b) Experimentally determined spatially resolved concentration at selected times. (c) Simulated spatially resolved concentration at the same times as in (b). }
\label{fig:evol-field-up}
\end{figure}

In all the experiments noted above, the gravitational force does not oppose the magnetic force. Instead, the gravitational force drives the particles downward, while the magnetic force also draws them downward and toward regions with a stronger magnetic field gradient. To examine the net effect of magnetic force and its interplay with the gravitational force in this system, the cuvette was moved to a lower position such that the liquid-air interface at the top of the cuvette is at the center of the magnetic field, and the simulation was modified accordingly. This has the same effect as inverting the magnetic field and gradient along the $y$ axis in simulations. In this configuration, the gravitational force competes with the magnetic force. Figure~(\ref{fig:evol-field-up}) shows the temporal evolution of the manganese oxide particle concentration for the same conditions as in Fig.~(\ref{fig:evol-field-down}) but with the new cuvette location, for which magnetic force and gravitational force compete. Although the temporal evolution of the spatially averaged concentration in the domain is nearly identical for the two field directions, the spatio-temporal evolution of particles in the domain are significantly different. In Fig.~\ref{fig:evol-field-up}(b), the competing forces of magnetophoresis and sedimentation induce a \modified{spatiotemporally complex} flow that is not observed in Fig.~\ref{fig:evol-field-down}(b) {(see Video 2 in the supplementary materials)}. {In contrast to the experiments shown in Fig.~\ref{fig:evol-field-down}(b), which exhibit the formation of well-developed vortices near the bottom of the cuvette, the experiments in Fig.~\ref{fig:evol-field-up}(b) do not display any clearly defined vortical structures, and are \modified{spatiotemporally complex} in nature.} Additionally, the depletion is initially greatest at the bottom of the cuvette, with the particles migrating upward. {The latter is expected because magnetic field gradients are high at the top of the cell.} Simulations show similar {trends in that there is a region of high concentrations at the top of the domain. However, some differences are evident. For example, the simulations predict a region of elevated particle concentration immediately near the bottom and along the centerline of the domain, which are not observed experimentally. As with the observations in Fig.~\ref{fig:evol-field-down}(c), this discrepancy is likely attributable to geometric simplifications in the simulation setup and the use of a three-size particle distribution in the model.}



\begin{figure}
\centering \includegraphics{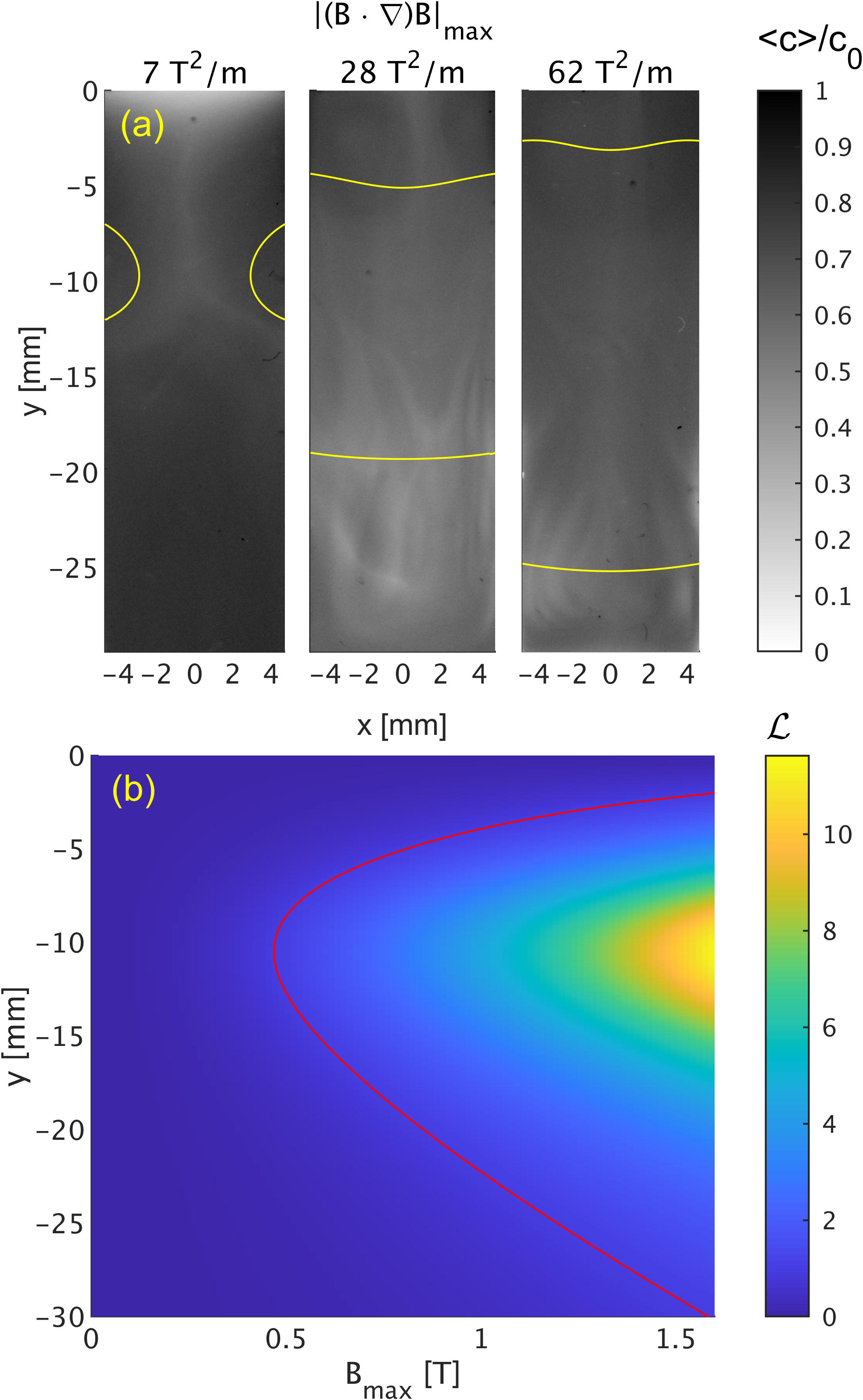}
\caption{Comparison of magnetophoresis with sedimentation using the parameter $\mathcal{L}$ defined in Eq.~\ref{eq:mag-vs-sed}. (a) Experimentally determined spatially resolved concentrations following exposure to magnetic field strengths for 3~h, 2~h, and~1~h respectively. Yellow lines outline the regions where $\mathcal{L} > 1$, and the maximum \BgradB~along the cuvette center line is listed above each image. (b) The value of $\mathcal{L}$ along the cuvette center line for different magnetic field strengths. The red line marks $\mathcal{L} = 1$.}
\label{fig:capture-region}
\end{figure}

In the arrangement shown in Fig.~\ref{fig:evol-field-up}, the magnetic gradient force and gravity act in opposite directions. To compare the magnitude of these forces, a parameter $\mathcal{L}$ is defined as the ratio of the magnetic and gravitational \Peclet numbers. Equations~\ref{eq:mag-Peclet} and \ref{eq:sed-Peclet} are combined to form:
\begin{equation}
{\mathcal{L} = \frac{\mathrm{Pe}_m}{\mathrm{Pe}_g} = \frac{\Delta \chi_V \left(\mathbf{B} \cdot \nabla \right) \mathbf{B}}{\mu_0 \Delta \rho g}}
\label{eq:mag-vs-sed}
\end{equation}
Equation~(\ref{eq:mag-vs-sed}) shows that the parameter $\mathcal{L}$ depends only on the magnetic field gradient and the physical properties of the particle material. Figure~\ref{fig:capture-region}(a) highlights the regions where $\mathcal{L} > 1$ as yellow overlays on the spatially resolved concentration of manganese oxide particles at intermediate field strengths with the maximum field at the top of the cuvette. The times shown are 3 h, 2 h, and 1 h respectively; these are the times where the effect of flow is clearest (compare with Fig.~\ref{fig:conc-evol}, where the standard deviation of the concentration reaches its steady value at these times). For the field at one-quarter strength (\BgradB~ = 7 T$^2$/m), $\mathcal{L} < 1$ throughout the cuvette except for two small regions near the pole face edges. The concentration is depleted outside these regions, though within the regions, the concentration is greater. For the half-strength field (\BgradB~ = 28 T$^2$/m), $\mathcal{L} > 1$ through a region spanning approximately half of the cuvette (between the two yellow curves). With the field at three-quarters strength (\BgradB~ = 62 T$^2$/m), $\mathcal{L} < 1$ is limited to the top 3 mm and bottom 5 mm of the cuvette. In Fig.~\ref{fig:capture-region}(b), $\mathcal{L}$ along the cuvette center line ($x = 0$) is shown as a function of field strength. The onset of $\mathcal{L} > 1$ is near 0.5~T, which suggests that magnetophoresis becomes dominant only above this imposed magnetic field. Additionally, the region where $\mathcal{L} > 1$ increases in size with increasing field strength until at 1.6~T, $\mathcal{L} > 1$ throughout the cuvette except for a small region near the field center ($y < 2$~mm where $\mathcal{L}<1$) indicating that sedimentation consistently dominates in this area. At the full field strength, $\mathcal{L} = 11.6$ is at its maximum, meaning magnetophoresis is a full order of magnitude stronger than sedimentation at this point.\par
\subsection{Particle size distribution and the possibility of field-induced aggregation}
Finally, we return to the phenomenon introduced earlier in the manuscript: field-induced particle cluster formation. The nanoparticles used in this study are polydisperse in size distribution, and Eq.~\ref{eq:total-flux} predicts greater flux for particles of larger radius. To reflect this size distribution, a three-species model was used for simulations, with all physical properties held constant except $R_p$. For each simulation, the particle sizes and relative abundances were tuned such that the simulated results best fit the experimental results. Table~\ref{tbl:sim-size-dist} shows the matching particle size distributions for Figs.~\ref{fig:evol-field-off} and \ref{fig:evol-field-down}, and the parameter $f$ is the mass fraction of each size of particle relative to the mixture of particles. For simulations both with and without the magnetic field applied, small particles (radius 80 nm) are most abundant, and larger particles are present in lower abundance. The latter may also be aggregates of smaller particles, which can be reduced but not fully eliminated through sonication. Most noteworthy is that the distribution shifts toward larger particles as the field is applied, which indicates that some field-induced aggregation may be occurring. Prior studies of field-induced aggregation suggest that two conditions are necessary. First, the magnetic forces between particles must be sufficient to overcome random thermal motions. This condition  can be quantified as the magnetic coupling parameter $\Gamma$. For paramagnetic particles, $\Gamma$ is given by~\cite{Leong2020Langmuir}:
\begin{equation}
\Gamma = \frac{\pi R_p^3 \Delta \chi_V^2 B^2}{9 \mu_0 k_B T}
\label{eq:mag-coupling-param}
\end{equation}
Secondly, {the formation of aggregates has been shown to be governed not only by magnetic energy but also by thermodynamic constraints~\cite{faraudo2016predicting}. In particular, while magnetic interactions promote aggregation, the associated loss of entropy acts as a counterforce. A thermodynamic framework developed before~\cite{Andreu2011aggregation} reveals that the competition between these effects is captured by a dimensionless aggregation parameter, $N^*$, which depends on both the magnetic properties of the particles and thermodynamic variables such as concentration and temperature, and is given by~\cite{faraudo2016predicting}:}
\begin{equation}
N^* = \sqrt{\frac{c_0}{\rho} \exp{(\Gamma - 1)}}
\end{equation}
Field-induced aggregation is only predicted for $\Gamma > 1$ and $N^* > 1$. Our calculations indicate that for particles smaller than 60 nm, $\Gamma < 1$, and field-induced aggregation is not predicted at any concentration (see Fig.~S3 in the supplementary materials). For $c_0 = 100 \mathrm{mg/l}$, field-induced aggregation is predicted only for $R_p > 130 \mathrm{nm}$. Therefore, the field in this study is not sufficiently strong to induce aggregation of nominal size nanoparticles; however, the larger particles and aggregates present may undergo field-induced aggregation. This is consistent with the results of numerical simulations that show slight increase in the largest size and abundance of the larger sized particles in Table~\ref{tbl:sim-size-dist}. It is important to note that the calculations of $\Gamma$ and $N^*$ are based on the assumption of a uniform magnetic field {and low overall particle concentration ($c_0/\phi < 0.1$)}. However, in our experiments and simulations, the magnetic field is strongly non-uniform, introducing additional magnetic forces that nanoparticles may experience. To the best of our knowledge, a theoretical analysis of field-induced cluster formation under non-uniform magnetic fields is not yet available. Nonetheless, we hypothesize that these additional forces associated with the non-uniform field further enhance field-induced cluster formation. As a result, the minimum particle radius capable of undergoing clustering is likely smaller than 130 nm under the conditions studied.\par 

{As discussed earlier, while the current transport model captures the essential features of particle migration under magnetic fields, certain discrepancies remain, particularly in the spatio-temporal evolution of concentration and flow profiles. The current mass transport model, even with multiple particle sizes, assumes a constant diffusion coefficient based on the Stokes–Einstein relation. However, this simplification overlooks hydrodynamic interactions between particles, which may become increasingly important at higher concentrations. In systems subjected to strong magnetic field gradients, magnetophoresis and field-induced aggregation can elevate the local particle concentration within micron-scale regions. Although our imaging resolution does not resolve concentration variations at this scale, it is well established that such microstructural crowding alters the local diffusivity of particles~\cite{leighton1987shear,qiu1990hydrodynamic}. Consequently, the diffusion coefficient may vary with both space and time, suggesting the potential relevance of a concentration-dependent, possibly tensorial diffusivity in capturing anisotropic transport behavior. Incorporating such effects in future modeling efforts could help improve agreement with experimental observations and provide deeper insights into the transport dynamics of paramagnetic nanoparticle suspensions under magnetic fields.}



\section{Conclusions}

In this study, we report rapid removal of weakly paramagnetic nanoparticles of manganese oxide from suspensions under nonuniform magnetic fields of an electromagnet using both experiments and numerical simulations. Our results can be summarized as follows:\par 

Experiments reveal that the rate of magnetophoresis experienced by nanoparticles is independent of their initial concentrations but strongly dependent on the magnetic field gradients. At low magnetic field gradients, the particle removal rate is slow; however, as the gradient increases, the removal rate rises significantly before plateauing for \BgradB $>$ 62 T$^2$/m. Notably, unlike previous studies by Leong et al.~\cite{Leong2015SoftMatter,Leong2020Langmuir}, we observed persistent concentration gradients throughout the particle removal process. We hypothesize that these concentration gradients are associated with the {magnetophoretically induced flows that can be further quantified using} magnetic Grashof number. The formation of particle concentration gradients induces bulk fluid motion, which mitigates spatial variations in concentration across the flow cell and is driven by the magnetic Grashof number. In this study, the magnetic Grashof number is on the order of 1-10, which are significantly lower than those reported in previous studies where $\mathrm{Gr}_m > 100$.  The smaller magnetic Grashof number values in our work suggest that the {induced} flows are considerably weaker compared to those observed in studies involving superparamagnetic nanoparticles\cite{Leong2020Langmuir,Leong2015SoftMatter}. Consequently, strong concentration gradients are expected and indeed observed in this study.\par 

Additionally, for configurations where magnetophoresis opposes sedimentation, we observe distinct behavior characterized by \modified{spatiotemporally complex} flows forming within the cuvette. Notably, well-defined regions emerge where magnetophoresis dominates over sedimentation; the extent of these regions can be quantified by comparing the magnetic and gravitational \Peclet numbers. Furthermore, our numerical simulations predict the formation of field-induced particle aggregation, consistent with previous theoretical analyses on field-induced cluster formation. These insights
highlight the potential of magnetic separation for sustainable metal recovery, offering a scalable
and environmental friendly solution for recycling critical materials from spent electronics.

\section{Acknowledgments}

A portion of this work was performed at the National High Magnetic Field Laboratory, which is supported by the National Science Foundation Cooperative Agreement No. DMR-1644779 and the state of Florida. This work was supported by the Center for Rare Earths, Critical Minerals, and Industrial Byproducts, through funding provided by the State of Florida. {We would like thank Dana Ezzeddine, Daniel Barzycki, and the Ricarte Lab for assistance with dynamic light scattering measurements and use of their instrument.}


\bibliography{refs}

\begin{thebibliography}{38}
\bibitem [{not()}]{note-deltachimol}%
  \BibitemOpen
  \href@noop {} {}\bibinfo {note} {$\Delta \chi_{m,i} = \chi_{m,i} - \frac{\mathcal{M}_i / \rho_i}{\mathcal{M}_l / \rho_l} \chi_{m,l}$, where $\chi_{m,i}$ is the molar magnetic susceptibility of component $i$ with no correction for the surrounding medium, $\mathcal{M}_i$ is the molar mass of component $i$, $\rho_i$ is the solid phase density of component $i$ (mass of material per unit volume of the solid material), and $\chi_{m,l}$, $\mathcal{M}_l$, and $\rho_l$ are these respective characteristics of the surrounding liquid medium.}\BibitemShut {Stop}%
\end{thebibliography}

\end{document}